\documentclass[a4paper]{jpconf}
\usepackage{graphicx}
\usepackage{subfigure}

\begin{document}

\title{Melting curve and Hugoniot of molybdenum up to 400 GPa by \emph{ab initio} simulations}

\author{C. Cazorla$^{1,2}$, M. J. Gillan$^{1,2}$, S. Taioli$^{3}$ and D. Alf\`e$^{1,2,3}$}
\address{$^{1}$London Centre for Nanotechnology, UCL, London WC1H OAH, U.K. \\ 
$^{2}$Department of Physics and Astronomy, UCL, London WC1E 6BT, U.K. \\
$^{3}$Department of Earth Sciences, UCL, London WC1E 6BT, U.K.}

\begin{abstract}
We report \emph{ab initio} calculations of the melting curve and Hugoniot of molybdenum for
the pressure range $0 - 400$~GPa, using density functional theory
(DFT) in the projector augmented wave (PAW) implementation. 
We use the ``reference coexistence'' technique to overcome
uncertainties inherent
in earlier DFT calculations of the melting curve of Mo. 
Our calculated melting curve
agrees well with experiment at ambient pressure and is consistent with shock
data at high pressure, but does not agree with the high pressure melting curve
from static compression experiments. Our calculated $P(V)$ and $T(P)$ Hugoniot relations
agree well with shock measurements. We use calculations of phonon dispersion relations
as a function of pressure to eliminate some possible interpretations of
the solid-solid phase transition observed in shock experiments on Mo.
\end{abstract}

\section{Introduction}
\label{sec:introduction}

The melting curves of transition metals at pressures
up to the megabar region are highly controversial,
particularly for b.c.c. metals.
Diamond anvil cell (DAC) measurements find that the melting temperature
$T_{\rm m}$ increases by only a few hundred K over the range
1 - 100 GPa~\cite{errandonea01,errandonea03}, while shock experiments indicate 
an increase of several thousand K over this range~\cite{brown83,hixson89,hixson92}.
There have been several \emph{ab initio} calculations on
transition-metal $T_{\rm m}$(P) curves, and the predictions
agree more closely with the shock data than with
the DAC data~\cite{moriarty94,moriarty02,wang02,belonoshko04}. A challenging
case is molybdenum, where there are very large
differences between DAC and shock data~\cite{errandonea05}, and where the
shock data reveal {\em two} transitions, the one at
high pressure ($\sim$ 380 GPa) being attributed to melting, and
the one at low pressure ($\sim$ 210 GPa) to a transition from
b.c.c. to an unidentified structure~\cite{hixson89}. We report here on
new \emph{ab initio} calculations of $T_{\rm m}$($P$) for Mo, and on the
$P(V)$ and $T(P)$ relations on the Hugoniot. We also report
preliminary information that may help in searching for
the unidentified high-P solid phase of Mo.

We use density functional theory 
(DFT), which gives very accurate predictions
for many properties of transition metals, including
Hugoniot curves~\cite{gillan06}. DFT molecular dynamics (m.d.) was
first used to study solid-liquid equilibrium 12 years ago~\cite{morris94}, and 
several different techniques are now available for using it
to calculate melting
curves. In such calculations, no
empirical model is used to describe the interactions between
the atoms, but instead the full electronic structure, and hence
the total energy and the forces on the atoms, is recalculated
at each time step. There have been earlier DFT calculations
on the melting of Mo, but the techniques used were prone to
superheating effects~\cite{moriarty94,belonoshko04}. In the present work, we use the so-called
``reference coexistence '' techniques~\cite{alfe02b,vocadlo04,alfe99}, which does not suffer from 
this problem.

Our work has several aims. First, we want to improve on the
reliability and accuracy of the predicted melting curve of
Mo obtained from DFT; second, we use DFT to predict the $P(V)$
and $T(P)$ relations on the shock Hugoniot; third, we want
to identify the unknown solid phase of Mo observed in shock
experiments. Our tests on the accuracy of DFT for Mo, and our
extensive calculations of the Mo melting curve will be
reported in detail elsewhere~\cite{cazorla07},
so we present only a summary here. However, our very recent
calculations on the shock Hugoniot will be presented in more
detail. These are important, because
temperature is very difficult to measure in shock experiments~\cite{yoo93}
and DFT gives a way of supplying what is missing in the shock
data. Our search
for the unidentified solid structure of Mo is at the
exploratory stage, but we present results for phonon frequencies
as a function of pressure, which allow us to rule out some possibilities.

In the following, we summarise
our tests on the accuracy of the DFT techniques
(Sec.~\ref{sec:techniques_and_tests}), and outline the reference coexistence technique. 
In Sec.~\ref{sec:results} we present our results for the
DFT melting curve and Hugoniot of Mo up to 400~GPa, and the study of the phonon frequencies.
Discussion and conclusions are in Sec.~\ref{sec:discussion}.

\begin{figure}[t]
\centerline{
        \includegraphics[width=0.7\linewidth]{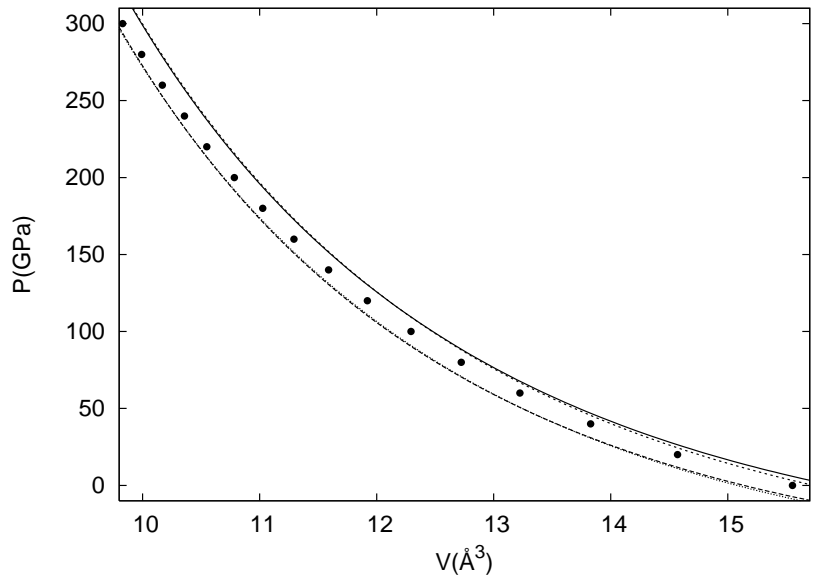}}%
        \caption{ Comparison between PAW and FP-LAPW results for the GGA(PBE) and 
                  LDA(CA) approximations for $E_{\rm xc}$. Solid and dashed curves show
		  GGA(PBE) and LDA(CA) FP-LAPW results, respectively; short-dashed
		  and dotted curves show GGA(PBE) and LDA(CA) PAW calculations,
		  respectively. Solid dots show experimental data~\cite{hixson92}.}
\label{fig:wien2kvasp}
\end{figure}

\section{Techniques and tests}
\label{sec:techniques_and_tests}

\begin{figure}[t]
\centerline{
        \includegraphics[width=0.7\linewidth]{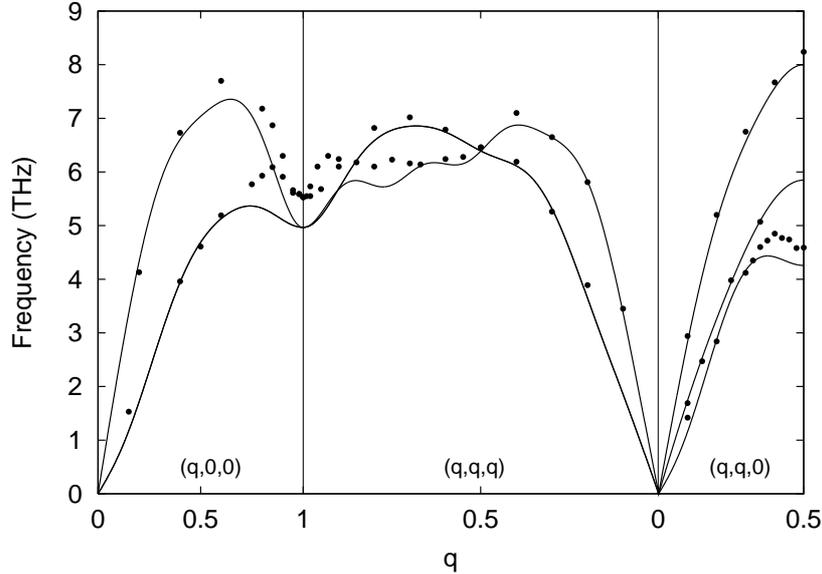}}%
        \caption{Comparison of calculated (curves) and experimental (dots)
	 phonon dispersion relations of Mo at zero pressure. 
	Experimental data are from Ref.~\cite{zarestky83}.}
\label{fig:vaspphonon}
\end{figure}

Our DFT calculations are performed mainly with the 
projector augmented wave (PAW) implementation~\cite{blochl94},
using the VASP code~\cite{kresse99,kresse96}, since PAW is
known to be very accurate. 
The main uncontrollable approximation in DFT is
the form adopted for the exchange-correlation 
functional $E_{\rm xc}$. To test the accuracy of PAW, and the
effect of $E_{\rm xc}$,
we have compared our predictions for the 
$P(V)$ relation of the b.c.c. Mo crystal against
low-$T$ experimental results (Fig.~\ref{fig:wien2kvasp}).
The pressure predictions from PAW using the
Perdew-Burke-Ernzerhof (PBE)~\cite{perdew96} and local-density
approximation (LDA)~\cite{ceperley80} forms of $E_{\rm xc}$ deviate
by $\sim 1.5$~\% in opposite directions from the experimental
data, but we adopt the PBE form,
because the deviations in this case are rather constant.
The PBE results of Fig.~\ref{fig:wien2kvasp} were obtained
with 4s states and below in the core and all other states
in the valence set. Inclusion of 4s states in the valence
set makes no appreciable difference to the PBE results.
We also tested the PAW implementation itself by repeating the
$P(V)$ calculations with the even more accurate full-potential
linearized augmented plane-wave (FP-LAPW) technique~\cite{lapw1,lapw2,lapw3}, using the WIEN2k code~\cite{wien2k}.
As shown in Fig.~\ref{fig:wien2kvasp}, PAW and
FP-LAPW results are almost identical. Further confirmation
for the accuracy of the PAW implementation and the PBE functional
comes from our comparisons of the calculated phonon dispersion
relations for the ambient-pressure b.c.c. crystal with experiment
(Fig.~\ref{fig:vaspphonon}).

The ``reference coexistence'' technique for calculating {\em ab initio}
melting curves has been described elsewhere~\cite{cazorla07}, but we recall
the main steps. First, an empirical reference model is
fitted to DFT m.d. simulations of the solid and liquid at thermodynamic
states close to the expected melting curve. Then, the reference model
is used to perform simulations on large systems in which solid
and liquid coexist, to obtain points on the melting curve of
the model. In crucial third stage, differences
between the reference and DFT total energy functions are used to
correct the melting properties of the reference model, to
obtain the {\em ab initio} melting curve. In the present work, the
total energy function 
of the reference model is represented by the 
embedded-atom model (EAM)~\cite{daw84,finnis84},
consisting of a repulsive inverse-power pair
potential, and an embedding term describing the d-band
bonding. The detailed procedure for fitting $U_{\rm ref}$ to DFT
m.d. data will be reported elsewhere~\cite{cazorla07}, but we note that we needed
to re-fit the model in different pressure ranges.

The simulations on solid and liquid Mo in stable thermodynamic
coexistence using the fitted reference model employed systems
of 6750 atoms, and we checked that the results do not change
if even larger systems are used. The protocols for preparing
these simulated systems, and for achieving and monitoring stable
coexistence were similar to those used in earlier work on the melting
curve of Cu~\cite{vocadlo04}. The procedures for correcting the melting curve of the
reference model, which depend on calculations of the the free energy
differences between the reference and DFT systems, have been described
and validated in earlier work~\cite{cazorla07}.

\section{Results}
\label{sec:results}

\subsection{Melting properties}
\label{sec:melting_properties}

\begin{figure}
\centerline{
        \includegraphics[width=0.7\linewidth]{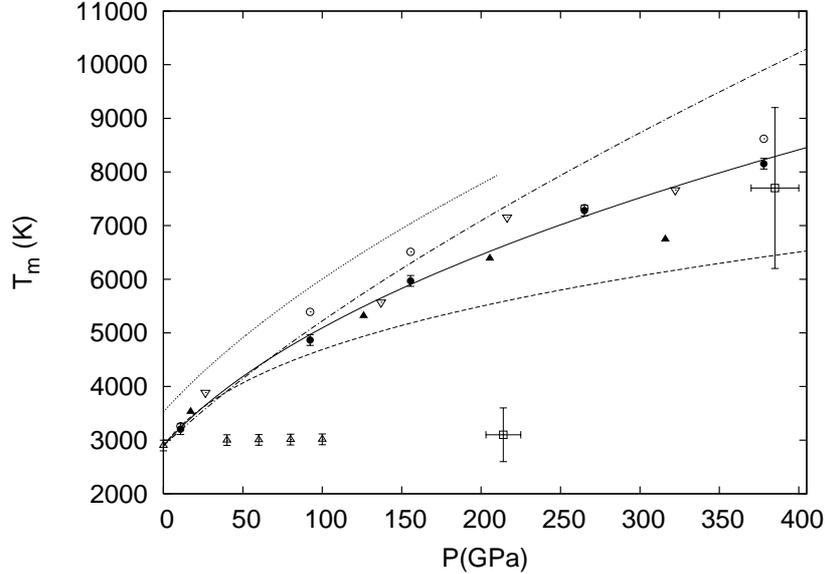}}%
        \caption{Calculated \emph{ab initio} melting curve (filled circles and solid line) of this work
                 compared with previous results: generalized pseudopotential calculations of 
		 Moriarty~\cite{moriarty94}(dotted line), dislocation-mediated models of 
                 Belonoshko \emph{et al.}~\cite{belonoshko04}(long-dashed line) and 
		 Verma \emph{et al.}~\cite{verma04}(dashed-dotted line); 
                 experimental shock-wave~\cite{hixson89} and DAC~\cite{errandonea01}
                 measurements are shown with empty squares and triangles, respectively. 
                 Filled and inverted-empty triangles show solid and liquid
                 \emph{ab initio} molecular dynamics calculations of 
                 Belonoshko \emph{et al.}~\cite{belonoshko04}, respectively.
		 Empty circles show results of this work obtained with the
		 EAM model without free-energy corrections.
                  }
\label{fig:melting}
\end{figure}

Fig.~\ref{fig:melting} shows the
melting curve of our reference EAM model, and the melting curve obtained
from this by correcting for the differences between the DFT and
reference total-energy functions; earlier {\em ab initio}-based
calculations of the Mo melting curve due to Moriarty and to
Belonoshko {\em et al.} are also indicated~\cite{moriarty94,belonoshko04}. We also
show points on the melting curve from DAC and shock
measurements~\cite{errandonea01,hixson89}. The differences between reference and
corrected melting curves are only a few hundred K, so that
the corrected curve should be very close to the melting
curve that would be obtained from the (PBE) exchange-correlation
functional if no statistical-mechanical approximations were made.
Because we avoid approximations of earlier {\em ab initio}-based
work, our present results should be a more accurate representation
of the DFT melting curve. Up to 100~GPa, the differences between our
results and earlier DFT work are rather small, and we 
confirm that DFT gives a much higher melting slope than that given
by DAC experiments. We obtain an accurate value of
$d T_{\rm m} / d P$ at $P = 0$ by fitting our melting
curve to the Simon equation $T_m = a ( 1 + P / b )^c$, with
$a = 2894$~K, $b = 37.2$~GPa, $c = 0.433$. The resulting $P = 0$
value of $T_{\rm m} = 2894$~K is close to the accepted experimental
value $T_{\rm m} = 2883$~K. Our $d T_{\rm m} / d P$ value
of $33.7$~K~GPa$^{-1}$ at $P = 0$ agrees with an older
experimental value of $33.3$~K~GPa$^{-1}$~\cite{shaner77}. Our DFT melting curve
is consistent with the point obtained at $P \simeq 370$~GPa from
shock measurements. However, we stress that the temperature of the
``experimental'' point was not measured, but estimated by considering 
superheating corrections to the shock-wave data~\cite{errandonea05,luo03}.  
Because of this, it is important to try and corroborate the estimated
experimental temperature, and this can be done by {\em ab initio}
calculations, as we explain next.

\begin{figure}
\centering
       { \includegraphics[width=0.47\linewidth]{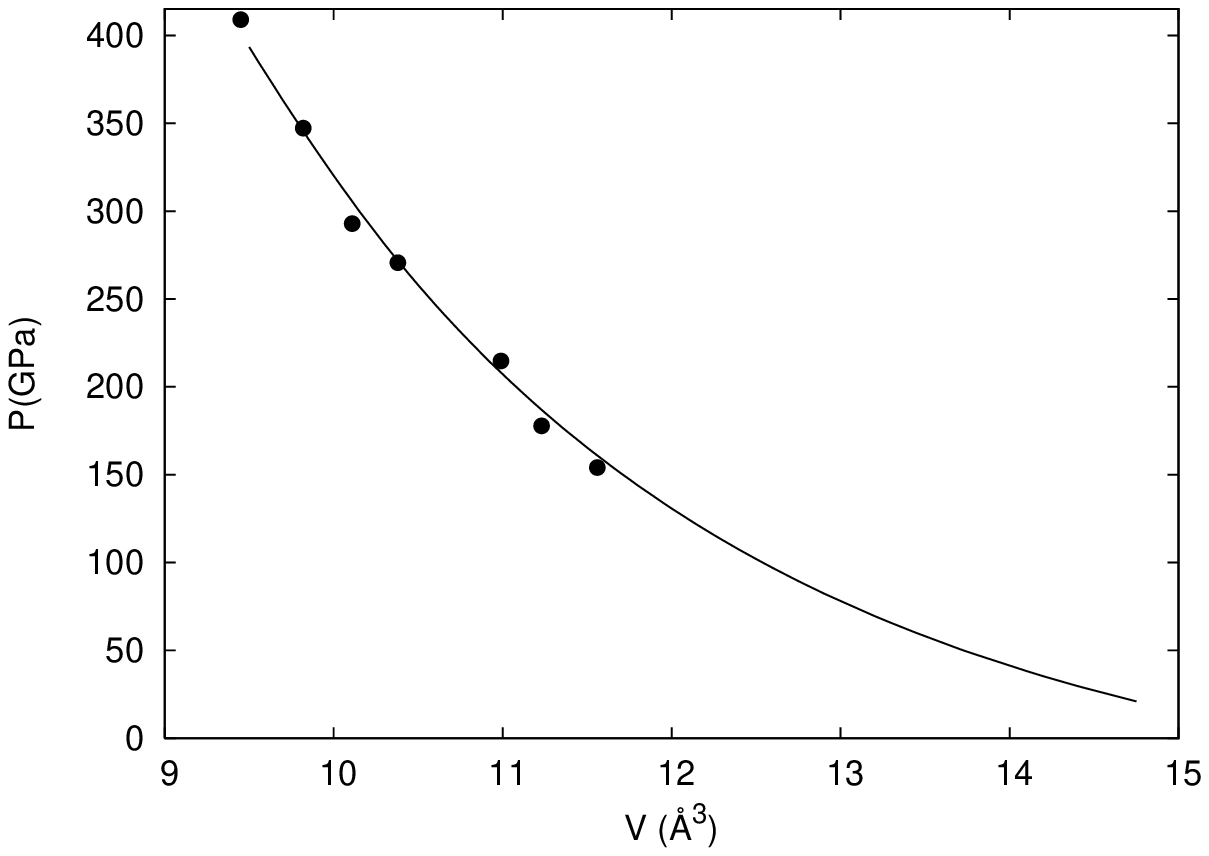} }%
       { \includegraphics[width=0.47\linewidth]{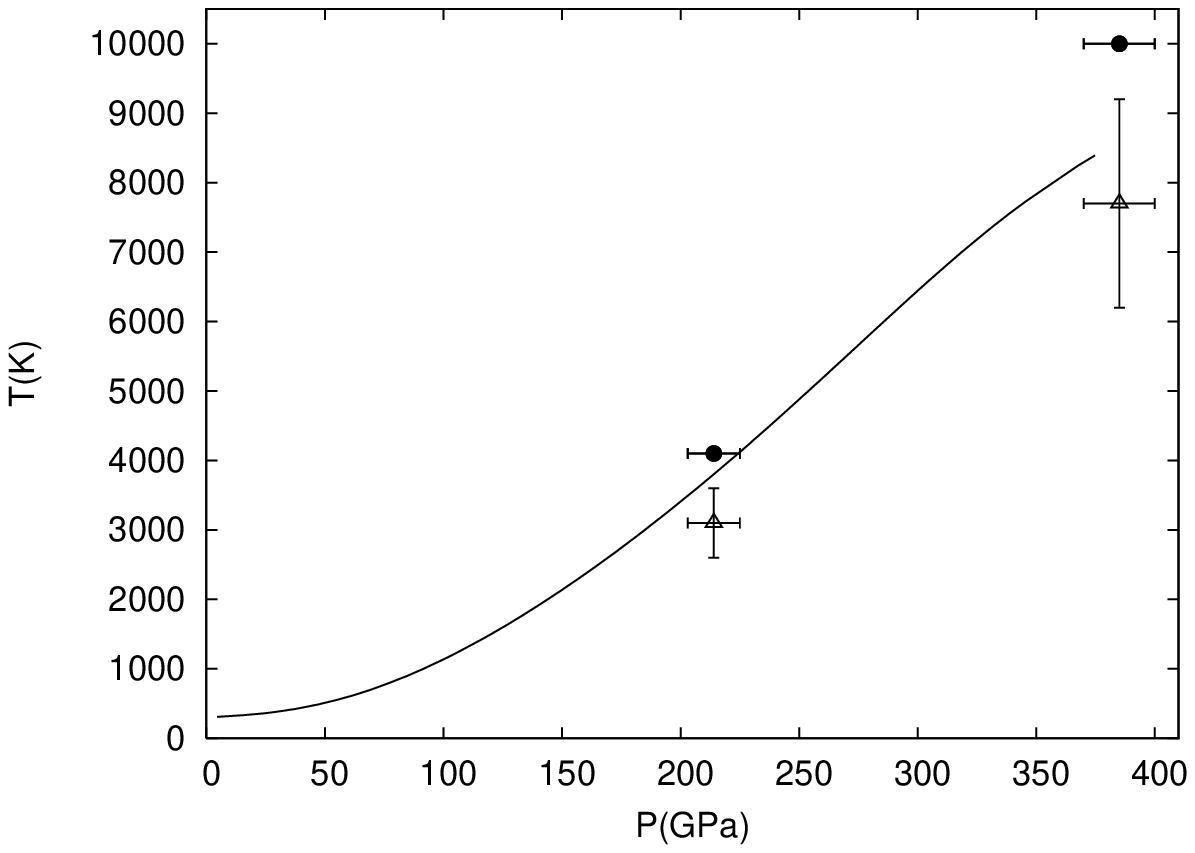} }%
\caption{{\it Left}: $P_{\rm H}(V_{\rm H})$ \emph{ab initio} relation on the Hugoniot of Mo up to pressure of 400 GPa;
	 dots reproduce experimental data of Refs.~\cite{hixson89,hixson92}. 
	 {\it Right}: $T(P_{\rm H})$ \emph{ab initio} relation on the Hugoniot of Mo up
	 to pressure of 400 GPa; dots reproduce experimental data of Refs.~\cite{hixson89,hixson92} and 
	 triangles the same results but corrected for superheating effects as given by Ref.~\cite{errandonea05}.
	 Error bars show the uncertainties in pressure and temperature.}
\label{fig:hugoniot}
\end{figure}

\subsection{Hugoniot curves}

The pressure $P_{\rm H}$, volume $V_{\rm H}$ and internal energy $E_{\rm H}$
in the shocked state are related to the initial volume $V_0$ and internal energy
$E_0$ by the well-known Rankine-Hugoniot formula~\cite{poirier91}:
\begin{equation}
\frac{1}{2}P_{\rm H}\left( V_{0} - V_{\rm H}\right) = E_{\rm H} - E_{0} \; .
\label{eq:hugo}
\end{equation}

Since the internal energy and pressure are given in terms of the Helmholtz
free energy $F$ by $E = F - T ( \partial F / \partial T )_V$ and
$P = - ( \partial F / \partial V )_T$, we can calculate the Hugoniot
from our DFT simulations, provided we can calculate $F$ as a function of
$V$ and $T$. So far, we have done this only for the b.c.c. crystal
in the harmonic approximation, in which 
$F ( V, T ) = F_{\rm perf} ( V, T ) + F_{\rm harm} ( V, T )$. Here,
$F_{\rm perf}$ is the free energy of the rigid perfect crystal, including
thermal electronic excitations, and $F_{\rm harm}$ is calculated from
the phonon frequencies $\omega_{{\bf q} s}$ (${\bf q}$ the wavevector, 
$s$ the branch). We calculate $F_{\rm harm}$ in the classical limit,
in which $F_{\rm harm} = 3 k_{\rm B} T \ln ( \beta \hbar \bar{\omega} )$
per atom, with $\beta = 1 / k_{\rm B} T$ and $\bar{\omega}$ is the geometric
average of phonon frequencies over the Brillouin zone. The methods
used for to calculate $F_{\rm perf} ( V, T )$ and the
frequencies $\omega_{{\bf q} s}$ were similar to those used in
our earlier work on Fe (Ref.~\cite{alfe01a}). For a set of
temperatures, we calculated $F_{\rm perf}$ at a set of
volumes, and fitted the volume dependence with a third-order
Birch-Murnaghan equation~\cite{birch78}. The temperature dependence of the
coefficients in this equation were then fitted with a
third-order polynomial. The phonon frequencies were calculated
at 12 volumes in the range $15.6 - 9.2$~\AA$^3$/atom, as
explained elsewhere~\cite{cazorla07}. The volume
dependence of the average $\bar{\omega}$ was then
fitted with a third-order polynomial.

To obtain $P_{\rm H} ( V_{\rm H} )$ and $T ( P_{\rm H} )$ from our
fitted free energy, for each value of $V_{\rm H}$ we seek the $T$
at which the Rankine-Hugoniot equation is satisfied, and from
this we obtain $P_{\rm H}$. For $V_0$ and $E_0$, we used values
from our GGA calculations; we checked that use of the experimental
$V_0$ made no significant difference. Comparison of our calculated
$P_{\rm H} ( V_{\rm H} )$ with measurements of Hixson 
{\em et al.}~\cite{hixson89,hixson92} (left panel of
Fig.~\ref{fig:hugoniot}) shows excellent agreement.
In the right panel, we compare our $T ( P_{\rm H} )$
with both uncorrected results of Hixson {\em et al.}
and also with results corrected for superheating, and
our results confirm their temperature estimates.

\begin{figure}
\centerline{
        \includegraphics[width=0.7\linewidth]{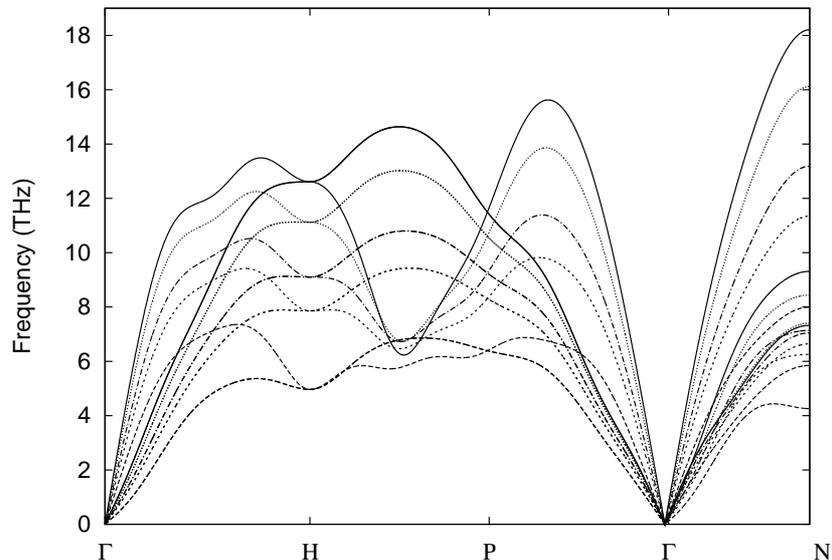}}%
        \caption{Comparison of the calculated phonon dispersion relations of Mo at  
	 different pressures: 0 GPa ($\broken$), 77 GPa ($\dashed$), 136 GPa ($\chain$), 274 GPa ($\dotted$) and 400 GPa ($\full$).}
\label{fig:freqcomp}
\end{figure}

\subsection{Solid-solid phase transition}

Efforts have been made to identify the high-$P$/high-$T$ structure of Mo 
indicated by shock experiments~\cite{hixson89}.
Hixson {\it et al.} used their theoretical prediction of b.c.c. $\to$ h.c.p. phase transition at 
$P \sim 320$~GPa and $T = 0$~K to suggest the h.c.p structure.
However, later calculations locate this transition at higher pressures
($420 \leq P \leq 620$ GPa), and temperature stabilisation of h.c.p. over b.c.c. below melting 
seems improbable. Furthermore, there are recent claims that under pressure 
b.c.c Mo transforms first to f.c.c. rather than h.c.p.~\cite{jona05}.
Very recently, Belonoshko {\it et al.}~\cite{belonoshko04} reported {\em ab initio} 
simulations on the f.c.c and A15 structures at
temperatures and pressures near those of interest (namely, $P \sim 210$~GPa and $T \sim 4000$~K),
and concluded that both structures are unlikely high-$T$ phases of Mo.  
We have performed {\em ab initio} calculations similar to those of Belonoshko {\it et al.} 
on the h.c.p. and $\omega$ structures ($P \sim 250$~GPa and $T = 5000$~K).
We find that temperature neither favours any of them respect to b.c.c., thus they must be excluded as well. 
Recently, a 2nd-order phase transition from cubic 
to rhombohedral has been 
observed in vanadium at $P = 69$~GPa~\cite{ding07}.      
This appears to be related to an earlier {\em ab initio} prediction of a phonon softening
in V~\cite{suzuki02}.
This suggests that a similar structural transition might occur in Mo.
We have used our calculated phonon dispersion relations of Mo over the range $0 - 400$~GPa
to test this. Fig.~\ref{fig:freqcomp} depicts our results at 0, 77, 136, 274 and 400~GPa,
but we see that no phonon anomaly like that reported for V is observed at any wavevector. This 
indicates that we should rule out structures based on small distortion of b.c.c.    

\section{Discussion}
\label{sec:discussion}

An important outcome of the present work is improved DFT calculations
of the melting curve of Mo over the pressure range $0 - 400$~GPa.
In particular, our techniques allow us to avoid the superheating 
errors which appear to affect an independent recent DFT study
of Mo melting~\cite{belonoshko04}. The accuracy of our calculations is confirmed
by the very close agreement with experiment for the melting
temperature $T_{\rm m}$ and the melting slope $d T_{\rm m} / d P$
at ambient pressure. The results fully confirm that the increase
of $T_{\rm m}$ by $\sim 2000$~K over the range $0 - 100$~GPa predicted
by DFT is about 10 greater than that deduced from DAC measurements.
A second important outcome is that our calculations of the temperature
along the shock Hugoniot support earlier temperature estimates
based on experimental data but corrected for possible superheating effects~\cite{errandonea05}.
This allows us to compare more confidently
the point on the melting curve at $P \simeq 380$~GPa derived from
shock data with our predicted melting curve, and we confirm that $T_{\rm m}$
at this pressure is {\em ca.} 8650~K. This is far above any reasonable
extrapolation of the DAC data. Concerning the search for the unknown
crystal structure of Mo indicated by shock experiments to exist above
{\em ca.} 220~GPa, we have been able so far only to rule out some
possibilities. Our calculations of the phonon dispersion realations
in the b.c.c. structure over the range $0 - 400$~GPa reveal no softening
of any phonons, and no indication of any elastic instability. This means
that the new crystal structure does not arise from small distortion of
b.c.c. This is interesting in the light of the recent discovery of
the elastic instability of b.c.c. V above $P \simeq 70$~GPa, predicted
initially by DFT, and observed very recently in x-ray diffraction experiments~\cite{ding07}. It seems
that the structural transition in Mo is of a different kind.

The large conflict between the melting curve of Mo derived from DAC
measurements on one side and from shock experiments and DFT
calculations on the other side must be due either to a misinterpretation
of the DAC data or to a combination of serious DFT errors and
misinterpretation of shock data. Given the accuracy of DFT that
we have been able to demonstrate (low-temperature $P(V)$ curve,
Hugoniot $P(V)$ curve, ambient-$P$ phonons), we think there is
little evidence for significant errors in DFT, which is also in
good accord with shock data. A possible explanation might be
that the large temperature gradients and non-hydrostatic stress
in DAC experiments might give rise to flow of material giving the
appearance of melting, even well below the thermodynamic
melting temperature. We also note recent evidence that temperature
measurement in DAC experiments may be subject to previously
unsuspected errors~\cite{benedetti07}, though probably not of the size needed to
resolve the conflict by themselves.

\ack
The work was supported by EPSRC grant EP/C534360, which is 50\% funded
by DSTL(MOD), and by NERC grant NE/C51889X/1. The work was conducted as part
of a EURYI scheme award to DA as provided by EPSRC (see www.esf.org/euryi).

\end{document}